\documentclass[a4paper]{PoS}
\pdfoutput=1
\usepackage{enumitem}

\title{Camera design and performance of the prototype Schwarzschild-Couder Telescope for the Cherenkov Telescope Array}

\ShortTitle{pSCT Camera Design and Performance}

\author{
C.~Adams~$ˆ{o}$ ,
G.~Ambrosi~$ˆ{l}$ , 
M.~Ambrosio~$ˆ{k}$~,
C.~Aramo~$ˆ{k}$~,
W.~Benbow~$ˆ{a}$ ,
B.~Bertucci~$ˆ{lm}$~,
E.~Bissaldi~$ˆ{h}$~,
M.~Bitossi~$ˆ{n}$~,
A.~Boiano~$ˆ{k}$~,
C.~Bonavolont\'a~$ˆ{k}$~,
R.~Bose~$ˆ{r}$ ,
A.~Brill~$ˆ{o}$ ,
J.~H.~Buckley~$ˆ{r}$ ,
M.~Caprai~$ˆ{l}$~,
L.~Di~Venere~$ˆ{i}$~,
Q.~Feng~$ˆ{b}$ ,
E.~Fiandrini~$ˆ{lm}$~,
N.~Giglietto~$ˆ{hi}$~,
F.~Giordano~$ˆ{hi}$~,
O.~Hervet~$ˆ{p}$,
G.~Hughes~$ˆ{a}$ ,
T.~B.~Humensky~$ˆ{o}$ ,
M.~Ionica~$ˆ{l}$~,
W.~Jin~$ˆ{c}$ ,
P.~Kaaret~$ˆ{e}$ ,
D.~Kieda~$ˆ{f}$ ,
B.~Kim~$ˆ{d}$ ,
F.~Licciulli~$ˆ{i}$~,
S.~Loporchio~$ˆ{hi}$~,
V.~Masone~$ˆ{k}$~,
T.~Meures~$ˆ{g}$ ,
B.~A.~W.~Mode~$ˆ{g}$ ,
R.~Mukherjee~$ˆ{b}$ ,
F.~R.~Pantaleo~$ˆ{hi}$~,
A.~Okumura~$ˆ{r}$~,
A.~Petrashyk~$ˆ{o}$ ,
J.~Powell~$ˆ{c}$ ,
R.~Paoletti~${nq}$~,
D.~Ribeiro~$ˆ{o}$ ,
A.~Rugliancich~$ˆ{nq}$~,
M.~Santander~$ˆ{c}$ ,
R.~Shang~$ˆ{d}$ ,
B.~Stevenson~$ˆ{d}$ ,
L.~Stiaccini~$ˆ{q}$~,
\speaker{L.~P.~Taylor}~$ˆ{g}$ ,
L.~Tosti~$ˆ{lm}$~,
V.~Vagelli~$ˆ{lm}$~,
M.~Valentino~$ˆ{jk}$
J.~Vandenbroucke~$ˆ{g}$ ,
V.~Vassiliev~$ˆ{d}$ ,
P.~Wilcox~$ˆ{e}$ ,
D.~A.~Williams~$ˆ{p}$,
for~the~CTA~SCT~Project \footnote{for consortium list see PoS(ICRC2019)1177}
\newline \\
\llap{$a$} Center for Astrophysics | Harvard \& Smithsonian, Cambridge, MA 02138, USA\\
\llap{$b$} Department of Physics and Astronomy Barnard College, Columbia University, NY 10027, USA\\
\llap{$c$} Department of Physics and Astronomy, University of Alabama, Tuscaloosa, AL 35487, USA\\
\llap{$ˆd$} Department of Physics and Astronomy, University of California, Los Angeles, CA, USA\\
\llap{$ˆe$} Department of Physics and Astronomy, University of Iowa, Iowa City, IA 52242, USA\\
\llap{$ˆf$} Department of Physics and Astronomy, University of Utah, Salt lake City, UT 84112, USA\\
\llap{$ˆg$} Department of Physics and Wisconsin IceCube Particle Astrophysics Center, University of Wisconsin, Madison, WI 53706, USA\\
\llap{$ˆh$} Dipartimento Interateneo di Fisica dell'Universit\'a e del Politecnico di Bari\\
\llap{$ˆi$} INFN Bari, Via E. Orabona 4, 70125 Bari, Italy\\
\llap{$ˆj$} CNR-ISASI, Italy\\
\llap{$ˆk$} INFN Napoli, Italy\\
\llap{$ˆl$} INFN Sezione di Perugia, Perugia, Italy\\
\llap{$ˆm$} Universit\'a degli Studi di Perugia, Perugia, Italy\\
\llap{$ˆn$} National Institute for Nuclear Physics (INFN), Pisa Section, L. B. Pontecorvo, 3 - 56127 Pisa, Italy\\
\llap{$ˆo$} Physics Department, Columbia University, New York, NY 10027, USA\\
\llap{$ˆp$} Santa Cruz Institute for Particle Physics and Department of Physics, University of California, Santa Cruz, CA 95064, USA\\
\llap{$ˆq$} Sezione di Fisica, Dipartimento SFTA dell' Universit\'a di Siena, Italy\\
\llap{$ˆr$} Institute for Space--Earth Environmental Research and Kobayashi--Maskawa Institute for the Origin of Particles and the Universe, Nagoya University, Nagoya 464-8601, Japan\\
\llap{$ˆs$} Department of Physics, Washington University, St. Louis, MO 63130, USA\\
E-mail: \email{ltaylor23@wisc.edu}}

\abstract{The Schwarzschild-Couder Telescope (SCT) is a candidate technology for a medium-sized telescope within the Cherenkov Telescope Array, the next generation ground based observatory for very high energy gamma ray astronomy. The SCT uses a novel two-mirror design and is expected to yield improvements in field of view and image resolution compared to traditional Cherenkov telescopes based on single-mirror-dish optics.
To match the improved optical resolution, challenging requirements of high channel count and density at low power consumption must be overcome by the camera. The prototype camera, currently commissioned and tested on the prototype SCT, has been developed based on millimeter scale SiPM pixels and a custom high density digitizer ASIC, TARGET, to provide 1600 pixels spanning a 2.7 degree field of view while being able to sample nanosecond photon pulses. It is mechanically designed to allow for an upgrade to 11,328 pixels covering a field of view of 8 degrees and demonstrating the full potential of the technology.  The camera was installed on the telescope in 2018.  We will present its design and performance including first light data.
}

\FullConference{36th International Cosmic Ray Conference -ICRC2019-\\
		July 24th - August 1st, 2019\\
		Madison, WI, U.S.A.}

\begin{document}

\section{Introduction}

Because Earth's atmosphere is opaque to very high energy (VHE) photons, ground based telescopes must detect them through indirect means. VHE gamma rays initiate air showers upon hitting the atmosphere. Shower constituents emit Cherenkov radiation which is then detected by imaging atmospheric Cherenkov telescopes (IACTs).

In order to detect gamma rays at energies of 50 TeV or higher, increased detection area is vital. The Cherenkov Telescope Array (CTA) is a ground-based observatory for very high energy gamma rays. The array will have a site in each hemisphere, will cover 4 $km^2$ in the south and 0.6 $km^2$ in the North, and will have telescope spacings of 100-300 m. \cite{acharyya2019monte} CTA will cover energies between 20 GeV and 300 TeV and have an order of magnitude greater sensitivity than current instruments. 

While most IACTs use single-mirror optics, the prototype Schwarzschild-Couder Telescope (pSCT) will use a novel dual-mirror design. The Schwarzschild-Couder optics produce an excellent optical point spread function, a wide field of view and a much smaller plate scale than traditional Davies-Cotton optics \cite{vassiliev2007schwarzschild}. Because of the small plate scale, traditional photomultiplier tubes can be replaced with silicon photomultipliers (SiPMs) \cite{otte2015development}. SiPMs are much smaller than traditional PMTs and so the pSCT has improved image resolution and thus better direction and energy resolution. The pSCT has 1600 pixels and a 2.68$^{\circ}$ field of view and is due to be upgraded to 11,328 pixels and an 8$^{\circ}$ field of view.

\section{Mechanical Design}

\begin{figure}
    \centering
    \includegraphics[width=0.5\textwidth]{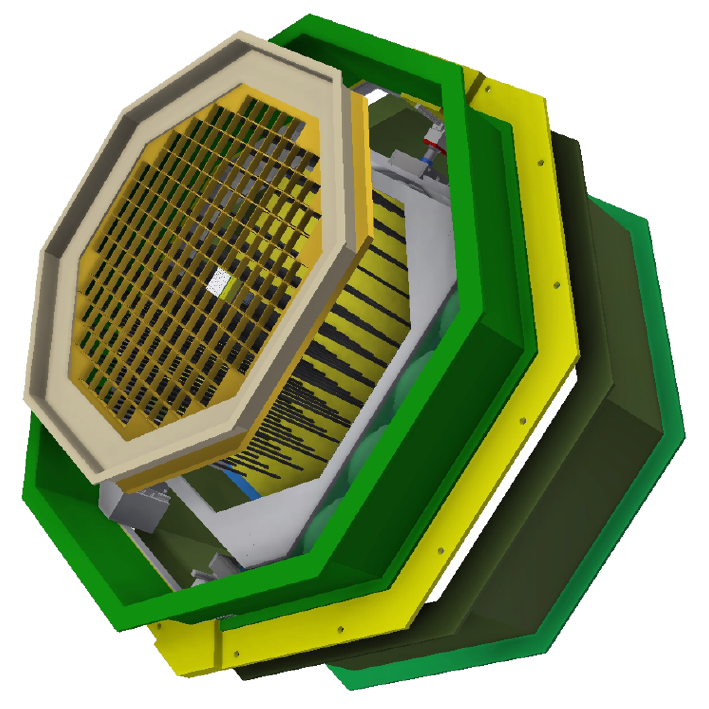}
    \includegraphics[width=0.1\textwidth, angle=90]{CameraModule.png}
    \caption{\textbf{Top}: exploded front view of the camera mechanical design. Modules are inserted through the front lattice and rest between the carbon fiber rods which connect the lattice to the back bulkhead. There are two holes per module in the back bulkhead. The first is for a backplane connector which connects the FEE primary board to the backplane electronics mounted to the back bulkhead. The second is for a screw which secures the module in place. Enclosing the camera is a shroud (dark green). The camera is secured to the telescope along the outer edge (light green). The interior of the camera can be moved using three motors. \textbf{Bottom}: exploded side view of a single module. To the left is the FPM including the SIPM, insulating foam and heatsink. The auxiliary (top) and primary (bottom) boards are shown outside of the aluminum FEE housing. The primary board is longer than the auxiliary board because of the backplane connector.}
    \label{fig:MechanicalDesign}
\end{figure}

The pSCT camera is comprised of an outer structure (connected rigidly to the telescope) and an inner structure (which can be moved for the purpose of alignment with three motors). The inner structure contains a front lattice, internal carbon fiber rods, and a back bulkhead. Modules are installed through the front lattice and connect to backend electronics through the back bulkhead. This bulkhead has two holes for each module, one for the module connector (which connects the module to its backplane) and one for the securing screw (which holds the module in place). Figure \ref{fig:MechanicalDesign} shows the mechanical design of the camera. A shutter is mounted to the front of the camera enclosure to shade the camera in daylight.

Camera electronics are kept cool using a chiller system, fans, and a Peltier element. Chilled water exits the chiller unit and is pumped to the right hand side of the camera. Fans blow across the chilled pipes circulating cold air through the camera. Each module is equipped with a Peltier element and heatsink which is cooled via circulating air. The water is then returned to the chiller to be cooled again. Currently only the central sector of the camera is populated with modules. A series of baffles are used to move chilled air through just this section of the camera.

\begin{figure}
    \centering
    \includegraphics[width=\textwidth]{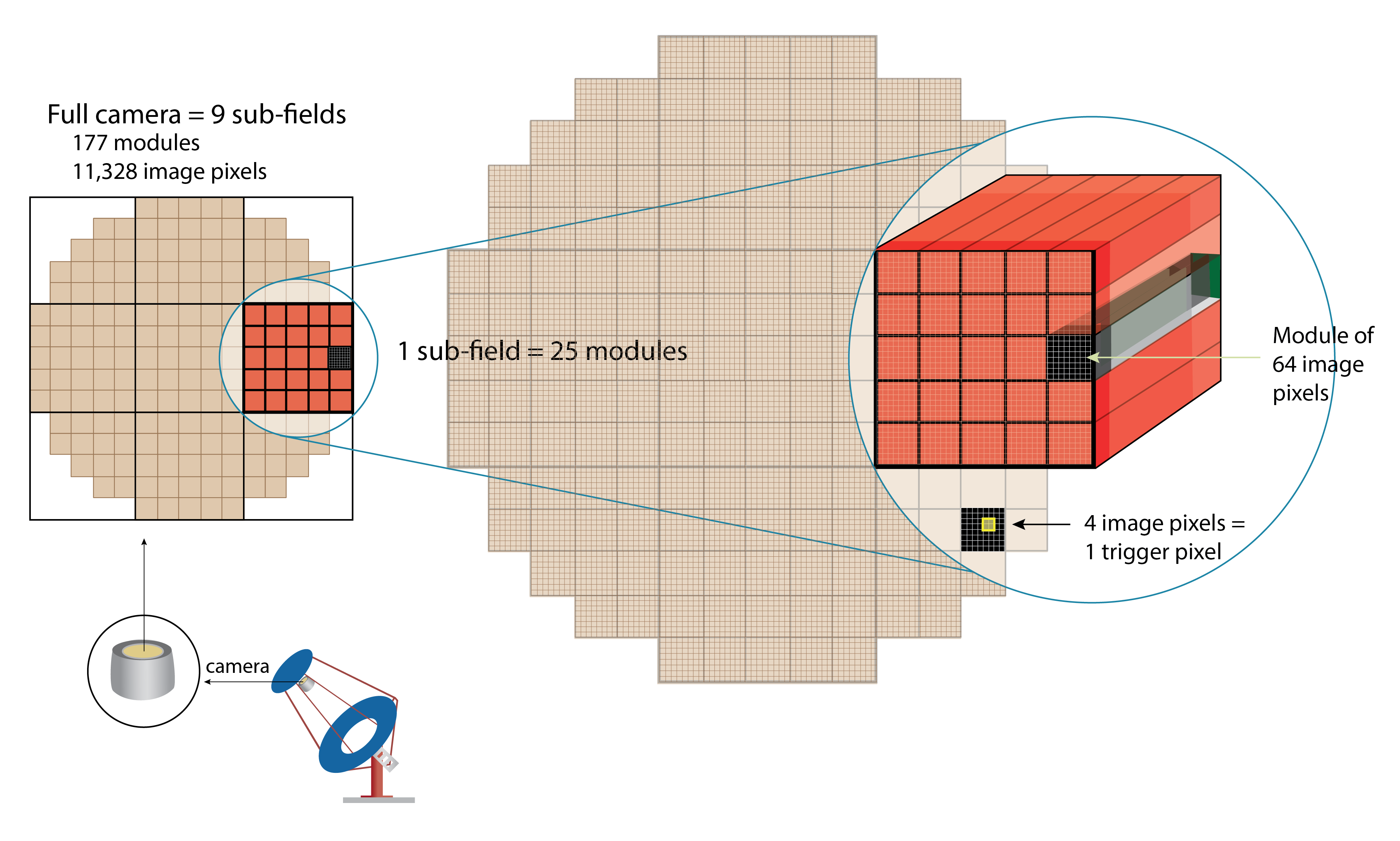}
    \caption{The pSCT camera has a hierarchical design. The full camera is comprised of 9 sectors, each of which can hold up to 25 camera modules. The sectors in each corner of the camera are equipped with fewer modules so that the entire camera will be roughly circular and have 177 modules. Each module has 64 image pixels so that the full camera will have 11,328 pixels. The current pSCT camera has only the center sector populated with modules. Currently 24 modules are installed in the central sector with the central module left uninstalled to accommodate optical alignment hardware.}
    \label{fig:CameraDesign}
\end{figure}

\section{Modules} \label{Modules}

Currently, 24 modules have been installed into the pSCT. Each module contains a Focal Plane Module (FPM) which sits on top of an aluminum module cage containing the Front End Electronics (FEE). Figure \ref{fig:MechanicalDesign} shows an exploded view of a camera module along with the camera enclosure which houses the modules. 

There are two types of FPMs currently used in the camera. The first is comprised of the commercially available Hamamatsu photon detector tiles S12642-0404PA-50(X). FPMs of this type are present on 15 of the 24 modules currently installed on the camera. The second is comprised of the third generation of Near Ultra Violet High Density SiPMs (NUV-HD3) produced by Fondazione Bruno Kessler (FBK) in collaboration with Istituto Nazionale di Fisica Nucleare (INFN) \cite{Luca}. FPMs of this type are present on the remaning 9 modules.

For each sensor type a tile has 16 SiPMs and a single module is equipped with 16 tiles in a square grid. The SiPMs are grouped such that each module has 4 quadrants of 16 image pixels each. Each quadrant is mounted on a printed circuit board which is then screwed onto the top of a copper post. Each quadrant is assigned a unique z-position such that the camera will approach a smoothly varying concave focal plane \cite{Otte:2015}. The copper post passes through a layer of foam and connects to a plastic baseplate. This baseplate is connected to the module FEE with screws. 

Each quadrant of the FPM is connected to the FEE with a micro-coaxial ribbon cable which provides power to the SiPMs and sends signals from the tile to the FEE. The Hamamatsu tiles have a median breakdown voltage of 64 V and will be operated at 3 V above breakdown. To achieve this, a uniform bias voltage (applied to all pixels in a sector) is combined with a trim voltage (adjustable for individual pixels). The trim voltages are set by the FEE and are actively adjusted.

The FEE is comprised of an auxiliary board and primary board stacked on top of one another within an aluminum housing. The boards are connected with high-speed connectors and the primary module connects directly to the backplane when mounted in the camera. 
The auxiliary board contains most of the analog processing. It includes a pulse shaping circuit for each pixel, current sensors for each group of 4 image pixels, and the trim voltage regulator. It also monitors the FPM temperature through a microcontroller and regulates the current through a Peltier element.

The primary board contains most of the digital control of the FEE. It includes triggering circuits for the SiPM signals, controls communications between devices in the FEE, and connects to the backplane and data acquisition (DACQ) boards. If requested by a backplane trigger, digitized data is transferred to the DACQ boards via Gigabit ethernet links. The primary board also contains four TARGET7 (7th generation TeV Array Readout with GSa/s sampling and Event Trigger) chips which sample, digitize and trigger on 16 channels \cite{funk2017target}. The data is sampled with a sampling array operating in "ping-pong" mode where one sampling group takes data for 32 ns while the other transfers data to a storage array. These units of 32 ns are called data blocks. The storage array is comprised of 512 blocks and can hold up to 16 $\mu$s of sampled data before having to overwrite. The FEE is also responsible for level one triggers. The analog sums of groups of four channels are monitored. If they cross a set threshold a trigger pulse is sent to the backplane for evaluation. When prompted by a backplane trigger, the storage buffer can be randomly accessed and transferred to a Wilkinson digitizer and then transferred to storage.

\section{Backend Electronics} \label{Backend Electronics}

As shown in Figure \ref{fig:CameraDesign}, the camera is divided into 9 sectors, each holding up to 25 modules and a single backplane PCB. Currently, only the central sector is populated with modules. The backplane for this sector is mounted onto the back bulkhead and connected to the 24 modules currently installed. DC-DC converters on the backplane convert the 70 V to the 12 V power required in the camera and modules. Each camera module connector includes pathways for high speed serial data, trigger signals, TAC messages required to localize triggering data, and 12 v DC power required for the module.

The backplane receives 16 level one triggers from each module through the primary board connector. The coincidence resolving time of the pattern trigger is determined by the pulse width of these trigger signals. A Trigger FPGA (TFPGA) on the backplane provides timing synchronization and trigger logic. The TFPGA forms coincidences from the 400 level one trigger inputs. The level two trigger requires that three level one triggers occur adjacent to one another. The backplane also performs housekeeping and power supply management. A second power supply module converts the 70 V to other voltages required by the backplane while a Housekeeping FPGA (HSKFPGA) is responsible for power-up sequencing for the various voltage levels. The HSKFPGA communicates with the TFPGA and Raspberri Pi to configure and monitor the trigger logic.

The backplane is mounted on the back bulkhead. Mounted on top of the backplane is a power board and power connectors as well as two DACQ boards each with two fiber cables leading to the network switch. Currently, there is no central module installed. Instead a positioner module is installed in its place which is designed to align the telescope optics. Also mounted on the back bulkhead is a fan control board and a Raspberry Pi which is connected to the backplane via an SPI connector. The Raspberry Pi is used to establish communication between the flashers and the backplane. This connection allows a user to control the DACQ board power, monitor the backplane power consumption and performance, and monitor modules (including supply voltages, currents, and synchronization messages). Additionally, software on the Pi allows a user to configure and trigger the three flashers at a selected rate and duration.

GPS time-tagging, timing synchronization, and trigger processing will be executed with the Distributed Intelligent Array Trigger (DIAT). The current pSCT has a 'demonstrator' version of the DIAT which covers three key features.

\begin{enumerate}[nolistsep]
    \item Provides a GPS-disciplined 62.5~MHz clock signal to up to 9 backplanes and actively synchronizes internal 125~MHz clock signals distributed to every FEE.
    \item Collects low-level trigger data from up to 9 backplanes to form a telescope trigger.
    \item Provides nanosecond accuracy GPS time-tagging for each telescope trigger.
\end{enumerate}
The 'demonstrator' DIAT is connected to the camera backplane via fiber optic cable and custom designed interface boards. Forced matching between the backplane 1 ns clock and the DIAT 1 ns time provides synchronization of the backend and front end electronics. 

The FEE does not have true 1 ns precision, rather it maintains an 8 ns clock (derived from the 125 MHz backplane signal) and uses the 64 bit 1 ns time to address the analog memory. SYNC messages generated by the DIAT allow the FEE to reset the counter to the next 8 ns edge.

The DIAT provides GPS time-tagging to each backplane and FEE. As a cross-check and backup, a stand-alone backplane event time tagging system has been installed on the pSCT, independent of the backplane. The system is comprised of a laser-diode module and a photo-detector/GPS module. The laser-diode module is located in the camera shroud and generates fast pulses at up to 5 kHz from camera backplane triggers. The photo-detector/GPS module is located in the electronics hut and is connected to the laser-diode via fiber optic communications cables. The laser pulses travel through these cables and arrive at the photo-diode. This output is then fed to a discriminator and a logic pulse is generated. A GPS time-stamp is attached to each pulse using a custom developed software and firmware and provides stable time-stamps for triggers to a GPS accuracy of better than 6 ns.

The event data generated by the camera is sent through the two DACQ boards to a data server (DS) located onsite. The server is responsible for data taking and general camera operation. It also acts as a local data repository. Both the slow control (responsible for powering and monitoring hardware components) and run control (responsible for loading configuration settings; starting, monitoring, and stopping runs; and recording data to disk) systems run on the DS.

\section{Auxiliary Systems} \label{Auxilliary Systems}

The pSCT LED flasher was developed for the CHEC-M camera \cite{Brown2015} and based on the design for the VERITAS flasher \cite{Hanna2010}. The LED Flasher consists of an LED board and a PSoC board linked by a 14-way Samtec SFMC connector. The LED board hosts ten 3 mm LEDs which provide light at 400$\pm$2.5 nm. The LEDs are behind a 20$^\circ$ circle pattern diffuser. Flasher triggers are received at the PSoC board where an active-high LED pulse is generated. The width of the pulse is determined by the setting of a potentiometer.

Three separate LED flashers are mounted on an optical table that is situated at the center of the secondary mirror. Each flasher has a unique potentiometer value (2.5 k$\Omega$, 1.8 k$\Omega$, 6 k$\Omega$) Each flasher is powered via an ethernet connection to the Raspberry Pi (see section \ref{Backend Electronics}) and receives TTL trigger pulses through a LEMO connector. Currently trigger pulses are provided by an Arduino located in the camera enclosure and controlled by the Raspberry Pi.

The current through individual LEDs is determined by the value of a resistor in series with that LED. Each flasher has 5 'bright' LEDs (each with a resistor of 80 $\Omega$) and 5 'dim' LEDs (with resistor values of 100, 110, 120, 140, and 130 $\Omega$). The light level of an LED depends both on the pulse width (set by the potentiometer) and the current (set by the resistor in series). Thus, each LED has a fixed output light level once installed on the telescope. Combination of LED on/off patterns can be used to achieve finer increments in light levels. Different patterns cover output light levels between approximately 8 and 80 photoelectrons (PEs). A single flasher covers a dynamic range just over two orders of magnitude. The full width at half maximum (FWHM) of all flasher pulses is well under 10 ns.

Power supplies and switches for the camera are located in two NEMA weatherproof cabinets mounted behind the camera enclosure. The camera fans are powered by an Acopian power supply. All other camera components are powered by a Wiener PL506 series power supply. This supply currently contains three power supply modules (with a total capacity of 6 modules), each providing an independent power channel. These modules (model PBN506-EX) are operated at 70 V with an 8 A capacity and $\sim$90\% efficiency. One module powers all 25 module SiPMs and is isolated from the others to reduce noise. The other two modules (tied together in parallel) power the module FEEs, backplane, and the Peltier system.


\section{Performance} \label{Performance}

\begin{figure}[t]
\centering
	\includegraphics[width=0.8\textwidth]{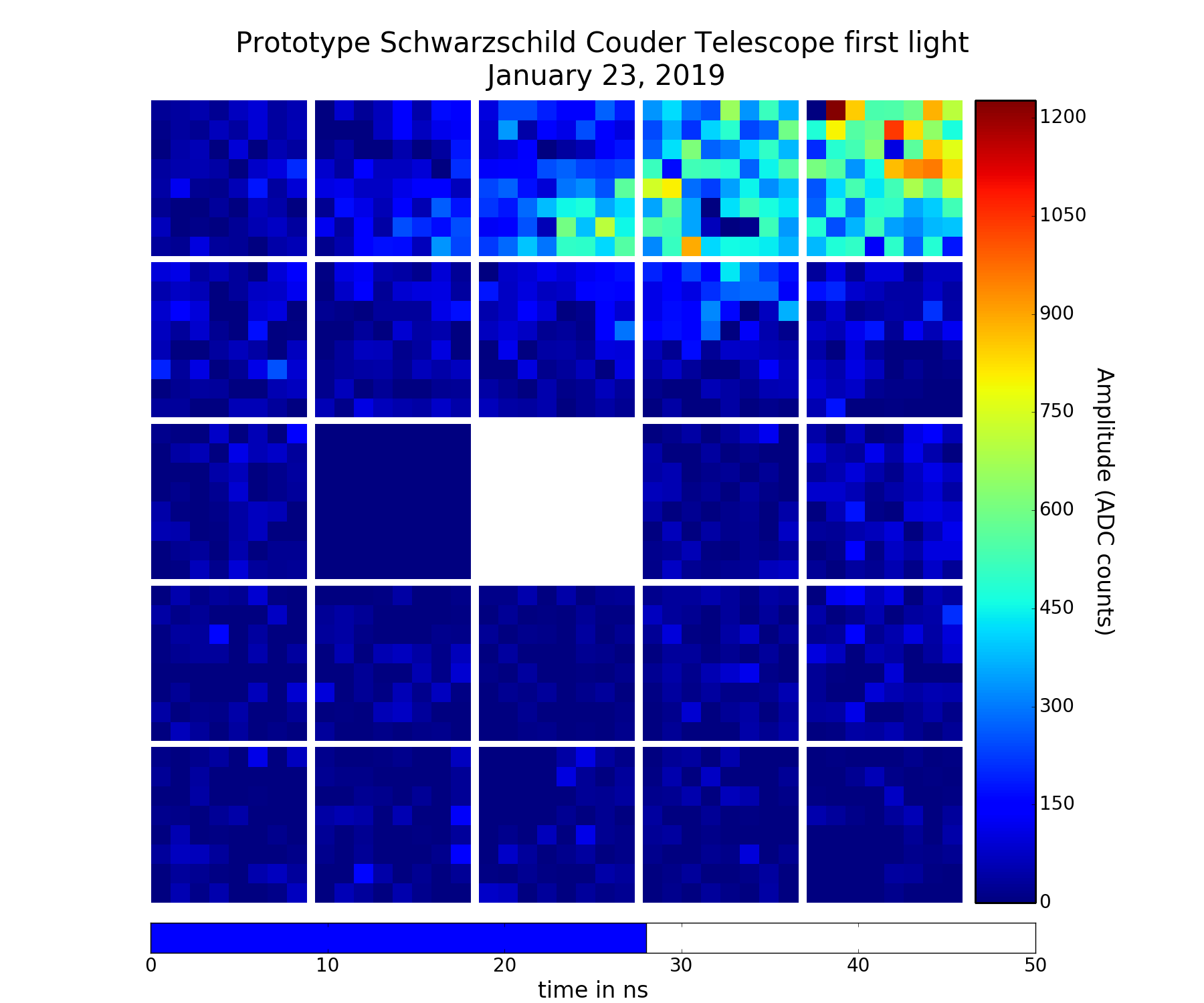}
	\caption[First Light]{Skyward pointing view of first light from the pSCT. First light and first shower candidates were achieved on January 23, 2019. This is a snapshot of a single event. This data has not yet undergone any pedestal subtraction or data processing. The module in the center left is dark because it failed to connect.}
	\label{fig:FirstLight}
\end{figure}

Inside the TARGET7 chip, the data path is subject to variations in physical routing and in the capacitance of the storage buffer elements. Because of this, information about which buffer elements have been used must be recorded with every waveform and a sample-dependent voltage to ADC count calibration must be applied. The current calibration strategy is comprised of pedestal subtraction and voltage to ADC conversion via a lookup table.

Pedestal subtraction requires that a database is created from pure electronic noise data to prevent bias from dark pulses. Dark pulses are pulses which occur in the SiPMs without any external light. This is achieved by disabling the SiPM bias voltage and instead using a software trigger to produce waveforms. These waveforms can then be averaged by storage array block and subtracted from the appropriate 32 ns sections of data. It is also important to note that the first block of a waveform shows a systematic elevation in ADC counts which must be accounted for in the calibration process.

The pSCT camera and optics were simultaneously uncovered for the first time in January 2019. Two data runs were taken, one with and one without simultaneous flashers, which contained both first light and first shower candidates. The optics were not fully aligned at the time of first light and thus some trace optical effects can be seen in the first light data (see Figure \ref{fig:FirstLight}). The optics are due to be fully aligned by the end of 2019.

\section{Summary and Outlook} \label{Summary and Outlook}

The pSCT camera will be upgraded over the next three years \cite{Thomas}. Modules will be upgraded to a lower noise design and the focal plane will be fully populated with modules. This will increase the number of pixels from 1600 to 11,328 and increase the FoV from 2.68$^{\circ}$ to 8$^{\circ}$. To accommodate these modules the number of backplanes will increase from 1 to 9. These modules will require synchronized timing between them and thus the full DIAT system will be implemented.

\section {Acknowledgements} \label{Acknowledgements}
This work is supported by National Science Foundation awards \#1229792, \#1828168, \#1707945, and a UW 2020 award from the University of Wisconsin. Further support comes from the agencies and organizations listed in \begin{footnotesize}  https://www.cta-observatory.org/consortium\_acknowledgments/ \end{footnotesize}. We thank these organizations for their financial contributions to the project.

\bibliographystyle{ICRC}
\bibliography{references.bib}

\end{document}